\documentclass[prd,nofootinbib,preprint,superscriptaddress,twocolumn,10pt]{revtex4}
\pdfoutput=1
\usepackage{amsmath,amssymb}
\usepackage{epsfig}
\usepackage{graphicx}
\usepackage[usenames,dvipsnames]{color}
\usepackage{subfigure}
\usepackage{slashed}
\usepackage{comment}
\usepackage[colorlinks,citecolor=blue]{hyperref}
\usepackage{color}

\usepackage{multirow}

\begin{document}

\title{Self Interacting Dark Matter and Dirac neutrinos via Lepton Quarticity }

	\author{Satyabrata Mahapatra}
	\affiliation{Department of Physics and Institute of Basic Science,
		Sungkyunkwan University, Suwon 16419, Korea}
	\author{Sujit Kumar Sahoo}
	\affiliation{Department of Physics, Indian Institute of Technology Hyderabad, Kandi, Sangareddy 502285, Telangana, India}			
	\author{Narendra Sahu}
	\affiliation{Department of Physics, Indian Institute of Technology Hyderabad, Kandi, Sangareddy 502285, Telangana, India}
	
	\author{Vicky Singh Thounaojam}
	\affiliation{Department of Physics, Indian Institute of Technology Hyderabad, Kandi, Sangareddy 502285, Telangana, India}
	\begin{abstract}

In this paper, we put forward a connection between the  self-interacting dark matter and the Dirac nature of neutrinos. Our exploration involves a $Z_4 \otimes Z_4'$ discrete symmetry, wherein the Dirac neutrino mass is produced through a type-I seesaw mechanism. This symmetry not only contributes to the generation of the Dirac neutrino mass but also facilitates the realization of self-interacting dark matter with a light mediator that can alleviate small-scale anomalies of the $\Lambda {\rm CDM}$ while being consistent with the latter at large scales, as suggested by astrophysical observations. Thus the stability of the DM and Dirac nature of neutrinos are shown to stem from the same underlying symmetry. The model also features additional relativistic degrees of freedom $\Delta N_{\rm eff}$ of either thermal or non-thermal origin, within the reach of cosmic microwave background (CMB) experiment providing a complementary probe in addition to the detection prospects of DM.

	
	\end{abstract}	
	\maketitle
	
\noindent
\section{Introduction}
Understanding the fundamental constituents of the Universe remains one of the paramount pursuits in modern physics. The intriguing interplay between dark matter (DM) and neutrinos has emerged as a focal point in this quest with the nature of neutrinos and dark matter standing out as mysteries. Although neutrino oscillation experiments offer valuable insights about neutrino masses and mixing \cite{deSalas:2020pgw,Super-Kamiokande:1998kpq,SNO:2001kpb,DoubleChooz:2011ymz,DayaBay:2012fng,RENO:2012mkc}, their findings are inconclusive in determining the intrinsic nature of neutrinos~\cite{ParticleDataGroup:2022pth}. Alternative experimental avenues such as neutrino-less double beta decay experiments ($0\nu\beta\beta$)~\cite{Auger:2012gs,KamLAND:2009zwo,Klapdor-Kleingrothaus:2000eir,IGEX:2002bce,Majorana:2013clq}, hold promise in establishing the Majorana nature of neutrinos. However, as of now, no such evidence exists, leaving the Majorana nature of light neutrinos unverified. This void has ignited a heightened interest in scrutinizing the plausibility of light Dirac neutrinos as a compelling alternative. 

Similarly the study of DM stands as a cornerstone in our quest to comprehend the underlying structure and dynamics of the Universe. Apart from the gravitational interactions, the nature and interactions of DM are still a mystery. Among the intriguing facets of DM, the concept of self-interacting dark matter (SIDM) has emerged as a compelling avenue that departs from the conventional paradigm of non-interacting dark matter particles. In contrast to its non-interacting counterpart, SIDM postulates large self-scattering of dark matter particles, providing a solution to challenges on smaller scales, such as the too-big-to-fail, missing satellite, and core-cusp problems - issues that collision less cold dark matter fails to address\cite{Spergel:1999mh,Tulin:2017ara,Peter:2012jh}.

Motivated by these, in this paper, we explore a flavor symmetric setup that can explain the origin of Dirac neutrino mass as well as give rise to a viable DM candidate with large self-interaction.  In particular, we focus on the realization of Dirac neutrino mass  through the influence of the cyclic symmetry $Z_4$. 
Such cyclic symmetry has already been considered in the literature as a discrete manifestation of the Lepton number symmetry and is quoted as {\it ''Lepton quarticity"}~\cite{CentellesChulia:2016rms,CentellesChulia:2016fxr,CentellesChulia:2017koy,Srivastava:2017sno}. Apart from this $Z_4$ symmetry, another $Z'_4$ symmetry is also imposed to forbid the direct tree-level coupling between left and right handed neutrinos as well as to realize a dark matter candidate.

After constructing the flavor symmetric model to incorporate the issues mentioned here, we focus on achieving the correct relic abundance of SIDM. There has been growing interest in the light DM regime, particularly in the GeV to sub-GeV scale, due to the null detection of DM at direct detection experiments~\cite{XENON:2023cxc,LZ:2022lsv}. However, coupling DM with light mediators to achieve  self-interaction cross-section to mass ratio
$\sigma/m \sim  1$ cm$^2/$g$\equiv  2 \times 10^{-24} $cm$^2/$GeV that can solve the small-scale anomalies often results in elevated DM annihilation rates leading to a relic abundance below the desired range for DM masses below a few GeV~\cite{Borah:2022ask}. Despite numerous proposed production mechanisms for SIDM, achieving the correct relic density remains challenging, though possible at the cost of introducing non-minimal aspects to the model~\cite{Borah:2022ask, Borah:2021yek, Borah:2021pet, Borah:2021rbx, Borah:2021qmi,Dutta:2022knf,Ghosh:2021wrk}. 

In addition, the Dirac nature of neutrino necessitates the existence of the right-handed neutrino $\nu_R$, which may contribute to the effective number of relativistic neutrino species, denoted as $N_{\rm eff}$ that can be probed by the CMB experiments~\cite{Planck:2018vyg,Abazajian:2019eic,SPT-3G:2019sok}. Present constraints from the CMB have imposed limitations ${\rm N_{eff}= 2.99^{+0.34}_{-0.33}}$ at the $2\sigma$ or 95\% confidence level. Anticipated future experiments such as CMB Stage IV (CMB-S4) \cite{Abazajian:2019eic} are poised to achieve unparalleled sensitivity, aiming to reach $\Delta {\rm N}_{\rm eff}={\rm N}_{\rm eff}-{\rm N}^{\rm SM}_{\rm eff} = 0.06$ at 2$\sigma$, thereby approaching the SM prediction. This heightened precision holds the potential to scrutinize our setup. We estimate the $\Delta N_{\rm eff}$ in our model depending
upon Yukawa couplings and masses of additional particles 
which can be of thermal and non-thermal origin and show the parameter space that can be probed by the future CMB experiments.  

The manuscript is built up as follows: In Section~\ref{sec::model}, we introduce our framework and discuss the Dirac neutrino mass generation in Section~\ref{sec::numass}. Then in Section~\ref{sec::neff}, we estimate the $\Delta N_{\rm eff}$ for our model outlining different possible thermal history for the $\nu_R$ production in the early Universe. Then we study the SIDM phenomenology in Section~\ref{sec::SIDM} discussing certain constraints in Section~\ref{sec::somecon}. We finally conclude in Section~\ref{sec::conclusion} and put several technical details in Appendices.




\section{The Model}\label{sec::model}
In this theoretical framework, the foundation is laid upon an underlying symmetry, where the Standard Model (SM) gauge group is extended with a discrete symmetry $Z_4 \otimes Z'_4$. To establish the genesis of Dirac neutrino mass, we introduce three right chiral fermions $\nu_R$s to the particle content, accompanied by three vector-like fermions $N(\equiv N_L+N_R)$. We introduce a singlet scalar $\eta$ to generate Yukawa coupling between $N$ and $\nu_R$.  
Additionally, we introduce another Dirac fermion $\chi$ as a potential DM candidate and a singlet scalar $S$ to mediate DM self-scattering due to its Yukawa coupling. The particle content and charge assignments under the imposed symmetry are detailed in Table~\ref{tab:particles}, where $\bold{z}$ and $\bold{z'}$ denote the fourth roots of unity, satisfying $\bold{z}^4 = 1$ and $\bold{z'}^4=1$. The {\it 'prime'} notation distinguishes charges under the two distinct cyclic symmetries imposed in this context. The charge assignments ensure that there is no direct coupling of $\nu_R$ with $\nu_L$.

The dual purpose of the imposed symmetry  is evident from Table~\ref{tab:particles}. Firstly, it prevents Majorana mass terms for $N$ and $\nu_R$. 
Secondly, it safeguards against the catastrophic couplings of potential DM candidates that challenge its stability. 
In particular, the incorporation of $Z'_4$ symmetry is crucial to secure the seesaw origin of neutrino mass, by preventing a tree-level coupling between left- and right-handed neutrinos and it also restricts certain DM couplings that could otherwise render DM unstable.

\begin{table}[h]
    \centering
\begin{tabular}{ | p{2cm}||p{2cm}||p{2cm} |  }
 \hline
 ~~~~~{\large Fields} & ~~~~~~{\large $Z_4$} & ~~~~~~{\large $Z'_{4}$} \\
 \hline
 ~~~~~~~$\overline{L}_L$ & ~~~~~~$\bold{z}^3$    & ~~~~~~$\bold{z'}^3$\\
 ~~~~~~~$l_R$ & ~~~~~~$\bold{z}$  & ~~~~~~1\\
 ~~~~~~~$\Phi$ & ~~~~~~1  & ~~~~~~$\bold{z'}$\\
 \hline
 \hline
 ~~~~~$N_{L,R}$ & $~~~~~~\bold{z}$ & ~~~~~~$\bold{z'}^2$\\
~~~~~~~$\nu_R$ & ~~~~~~$\bold{z}$  & ~~~~~~1\\
 ~~~~~~~$\eta$ & ~~~~~~1  & ~~~~~~$\bold{z'}^2$\\
 \hline
 \hline
 ~~~~~~~$\chi$ & ~~~~~~$\bold{z}$  & ~~~~~~$\bold{z'}$\\
 ~~~~~~~$S$ & ~~~~~~$\bold{z}^2$  & ~~~~~~$\bold{z'}^2$\\
 \hline
\end{tabular}\label{tab:particles}
    \caption{The charge assignment of the SM and BSM particles under the extended discrete symmetry. }
    \label{tab:particles}
\end{table}

The Lagrangian for the model dictated by the imposed symmetry is given by:
\begin{equation}
\begin{split}
        -\mathcal{L} \supset & \; f_{ij} \overline{L}_{L_i} \Tilde{\Phi}N_{R_j} +g_{ij} \overline{N}_{L_i} \eta \nu_{R_j} + M_{ij} \overline{N}_{L_i} N_{R_j}\\
        &  + y_\chi \overline{\chi^c}\chi S + m_\chi \overline{\chi}\chi ~-~ V(\Phi, \eta, S)\,,
\end{split}
\end{equation}
with $i,j=1,2,3$. 
The scalar potential $V(\Phi, \eta, S)$ can be written as:
\begin{equation}\label{eqn:potential}
\begin{split}
    V(\Phi, \eta, S)= & -\mu_h^2 (\Phi^\dagger \Phi) + \lambda_h(\Phi^\dagger \Phi)^2 \\
    & -\frac{\mu_\eta^2}{2} \eta^2 + \frac{\lambda_\eta}{4}\eta^4 +\frac{\lambda_{h \eta}}{2}(\Phi^\dagger \Phi)\eta^2 \\
    & +\frac{\mu_S^2}{2} S^2 + \frac{\lambda_S}{4}S^4 +\frac{\lambda_{h S}}{2}(\Phi^\dagger \Phi)S^2\\
    & +\frac{\lambda_{\eta S}}{4}\eta^2 S^2\,. 
\end{split}
\end{equation}


Here we assume that the fields $\eta$ and $S$ are real, a choice consistent with the real charges assigned to these fields ($\bold{z}^2=\bold{z'}^2=-1$). Before proceeding further, here it is worth noticing that, in the absence of terms involving $S^2$ and $S^4$, the theory exhibits an enhanced abelian global symmetry. This symmetry can be interpreted as a generalized global lepton number symmetry $U(1)_{\rm L}$, often invoked in the context of Dirac seesaw realizations for light neutrino masses. However, the presence of the $S^2$ and $S^4$ terms explicitly break this $U(1)_{\rm L}$ invariance, while the $Z_4 \otimes Z'_4$ remains the remnant symmetry group governing all interactions and still providing a viable framework for the realization of Dirac neutrino mass and self-interacting dark matter.

Since the SM Higgs doublet $\Phi$ and $\eta$ bear trivial charges under $Z_4$ but possess nontrivial $Z'_4$ charges, the acquisition of vacuum expectation values(vev) by $\Phi$ and $\eta$ leads to the spontaneous breaking of $Z'_4$ down to a remnant $Z_2$ symmetry, while the $Z_4$ symmetry remains unbroken and unaffected. Moreover, under the remnant $Z_2$ symmetry, $\chi$ is odd, while all the other particles are even. As a result, $\chi$ acts as a stable dark matter. Following the breaking of $Z'_4$ symmetry, neutrinos can subsequently attain a tiny non-zero mass through the Type-I Dirac seesaw mechanism, as illustrated in Fig.~\ref{fig:dirac_neutrino}, which we will discuss in details, in  section~\ref{sec::numass}.

The scalar $S$ being charged under $Z_4$ does not acquire any vev  thus ensuring that the $Z_4$ symmetry remains
unbroken. On the other hand, as $\eta$ and $\Phi$ both acquire vevs, a mixing occurs between $\eta$ and the Higgs field $h$. Parameterizing $\Phi$ and $\eta$ as 
\begin{equation*}
    \langle \Phi \rangle=\frac{1}{\sqrt{2}}
    \begin{pmatrix}
    0\\
    h'+v
    \end{pmatrix}  \; , \; \langle \eta \rangle=u+\eta'\,,
\end{equation*}
after the spontaneous symmetry breaking, the minimization conditions read as:
\begin{equation}
\begin{split}
    \mu_h^2= & v^2\,\lambda_h + \frac{u^2\, \lambda_{h \eta}}{2} \\
    \mu_\eta^2= & u^2\, \lambda_\eta + \frac{v^2\, \lambda_{h \eta}}{2}\,.
\end{split}
\end{equation}

Thus the scalar mass-matrix in the basis $(h'~~\eta'~~S)^T$ can be written as:
\begin{equation}
    \mathcal{M}=
    \begin{pmatrix}
        2v^2 \lambda_{h} & uv \lambda_{h\eta} & 0 \\
        uv \lambda_{h\eta} & 2u^2 \lambda_{\eta} & 0\\
        0 & 0 & M_S^2
    \end{pmatrix}
\end{equation}

Notably, the mass of $S$ remains unchanged as it does not mix with the other fields, but it does receive mass contributions from the vevs of $\Phi$ and $\eta$. After diagonalizing the resulting mass matrix, we obtain the physical states $h$ and $\eta$ as a linear combination of $h'$ and $\eta '$, with the masses of the particles $h$, $\eta$ and $S$ given by

\begin{eqnarray}\label{eqn::mass2}
  m_h^2 &\approx & 2v^2\,\lambda_h \\
    M_\eta^2 &\approx& u^2 \lambda_{\eta} ~ - \frac{u^2 }{2 \lambda_h} \left( \lambda_{h\eta}^2-2 \lambda_h \lambda_\eta \right)\\
    M_S^{2} &=& \mu_s^2+\frac{v^2\, \lambda_{h s}}{2} + \frac{u^2\, \lambda_{\eta s}}{2} 
\end{eqnarray}

Here we have used the approximation that $u \ll v$. 
The various parameters entering the Lagrangian can be expressed in terms of physical masses and the mixing angle as 

\begin{equation}
\begin{split}
    \lambda_h \, =& \frac{\,m_h^2\,\cos^2\theta + \,M_\eta^2\,\sin^2\theta}{2 v^2}\\
    \lambda_\eta \, =& \frac{M_\eta^2\,\cos^2\theta + m_h^2 \, \sin^2\theta}{2 u^2} \\
    \lambda_{h \eta} =& \frac{\sin2\theta (m_h^2 -M_\eta^2)}{2\,u\,v}
\end{split}
\end{equation}


\section{Dirac Neutrino}\label{sec::numass}

\begin{figure}[h]
    \centering
    \includegraphics[scale=1]{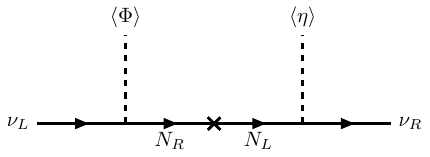}
    \caption{Dirac neutrino mass via type-I seesaw mechanism}
    \label{fig:dirac_neutrino}
\end{figure}

As outlined in the preceding section, the $Z_4$ charge assigned to $\nu_R$ effectively prohibits the Majorana mass term ($m\overline{\nu_R^c}\nu_R$), thereby establishing the viability of neutrino Dirac mass generation within our scenario. It is crucial to reiterate that this implementation allows $Z_4$ to be interpreted as a minimal discrete realization of the Lepton number symmetry $U(1)_L$. The Feynman diagram illustrating neutrino mass generation is depicted in Fig.~\ref{fig:dirac_neutrino}. This process is facilitated by a Dirac fermion $N$ and a singlet scalar $\eta$, where the latter breaks the $Z'_4$ symmetry while preserving the $Z_4$ symmetry, given its trivial charge under $Z_4$. It is noteworthy that the $Z_4$ charged scalar $S$ does not acquire a vacuum expectation value, ensuring the preservation of the $Z_4$ symmetry even after spontaneous symmetry breaking.

Consequently, the mass matrix for the neutrinos and $N$  can be written in the basis of $(\overline{\nu_L} ~~ \overline{N_L})$ and $(\nu_R ~~ N_R)^T$ as:
\begin{equation}
    M_{\nu N}=
    \begin{pmatrix}
        0 & \bold{m} \\
        \bold{m'} & M_N 
    \end{pmatrix}
    \label{eqn::6by6massmatrix}
\end{equation}

Here, the matrices $\bold{m}=\textbf{f}~v/\sqrt{2}$ and $\bold{m'}=\textbf{g}~u$ are both $3 \times 3$ matrices. 
Diagonalizing the above mass matrix we obtain the light neutrino mass matrix as:
\begin{equation}\label{eqn::neutrino mass}
    M_\nu = \bold{m}~ M_N^{-1}~ \bold{m'}
\end{equation}
The detailed calculation for this block diagonalization process is provided in Appendix \ref{app::neutrino mass}.

Diagonalizing the neutrino mass matrix (Eq.~\ref{eqn::neutrino mass}), we get the neutrino mass eigenvalues which should align with the neutrino oscillation data\cite{deSalas:2020pgw, ParticleDataGroup:2022pth}. To ensure this compatibility, we adopt a strategy akin to the Casas-Ibarra parametrization for Majorana neutrino masses \cite{Casas:2001sr}. 
We can diagonalize the above neutrino mass matrix (Eq.~\ref{eqn::neutrino mass}) by a bi-unitary transformation as: 
\begin{eqnarray}\label{eq::nubegin}
        \hat{M_\nu} &=& V_{lL}^\dagger~ M_\nu ~V_{lR} \nonumber\\
\sqrt{\hat{M_\nu}}\sqrt{\hat{M_\nu}}&=& V_{lL}^\dagger ~\bold{m} ~M_N^{-1} ~\bold{m'}~ V_{lR}
\end{eqnarray}

Here, without loss of generality, we assume $M_N$ to be diagonal and replace $M_N$  by its diagonal form $\hat{M_N}$ in the following discussion.
$V_{lL}$ corresponds to the transformation matrix for the left-handed neutrinos, typically the PMNS matrix without the Majorana phases. 
{\begin{widetext}
\begin{equation}
    V_{lL}=
    \begin{pmatrix}
        c_{12}c_{13} & s_{12}c_{13} & s_{13}e^{-i\delta} \\
        -s_{12}c_{23}-c_{12}s_{23}s_{13}e^{i\delta} & c_{12}c_{23}-s_{12}s_{23}s_{13}e^{i\delta} & s_{23}c_{13} \\
        s_{12}s_{23}-c_{12}c_{23}s_{13}e^{i\delta} & -c_{12}s_{23}-s_{12}c_{23}s_{13}e^{i\delta} & c_{23}c_{13} \\
    \end{pmatrix}
\end{equation}
\end{widetext}
}

where $c_{ij}=\cos(\theta_{ij})$ and $s_{ij}=\sin(\theta_{ij})$ for $i,j$ running from $1$ to $3$ and $\delta$ is the Dirac CP phase. The values of the oscillation parameters are given by \cite{deSalas:2020pgw}, $\theta_{12}=31.5^o- 38.0^o$, $\theta_{23}=41.8^o-50.7^o$, $\theta_{13}=8.0^o-8.9^o$ and $\delta=157^o-349^o$ at $3\sigma$ C.L. On the other hand, $V_{lR}$ represents the transformation matrix for the right-handed neutrinos. 
Multiplying on both sides of the (Eq.~\ref{eq::nubegin})  by $\sqrt{\hat{M_\nu}^{-1}}$, and rewriting the $\hat{M_N}^{-1}$ as the product of two square roots, we obtain
\begin{eqnarray}\label{eq::numid}
  I &=& \sqrt{\hat{M_\nu}^{-1}} V_{lL}^\dagger \bold{m} M_N^{-1} V_{lR} \bold{m'} \sqrt{\hat{M_\nu}^{-1}} \nonumber\\
        I &=& \left(\sqrt{\hat{M_\nu}^{-1}} V_{lL}^\dagger \bold{m} \sqrt{\hat{M_N}^{-1}} \right) \left(\sqrt{\hat{M_N}^{-1}} \bold{m'} V_{lR} \sqrt{\hat{M_\nu}^{-1}} \right)\nonumber\\
        &\equiv& R R^{-1} 
\end{eqnarray}
 In this context, $R$ represents a general complex matrix, contrary to the orthogonal matrix utilized in the Casas-Ibarra parametrization \cite{Casas:2001sr} for the complex symmetric Majorana neutrino mass matrix. 

From Eq.~\ref{eq::numid}, one can deduce the Yukawa couplings $\textbf{f}$ and $\textbf{g}$ as :
\begin{equation}\label{eq::fg}
\begin{split}
    \textbf{f} =& \frac{\sqrt{2}}{v}\left(V_{lL}\sqrt{\hat{M}_\nu}~R~\sqrt{\hat{M}_N}\right) \\
    \textbf{g} =& \frac{1}{u}\left(\sqrt{\hat{M}_N}~R^{-1}~\sqrt{\hat{M}_\nu}V_{lR}^\dagger\right).
\end{split}
\end{equation}


The matrix $R$ is a general complex matrix with 8 independent parameters. It plays a crucial role in tuning the couplings $\textbf{f}$ and $\textbf{g}$. We have considered the $R$ matrix to be diagonal, and we varied the diagonal elements in the range $[10^{-4}, 10^{7}]$. For simplicity, we assume the $V_{lR}$ matrix to be identity matrix. Thus $\mathbf{g}$ is a diagonal matrix whereas $\mathbf{f}$ has a general structure that explains the neutrino oscillation data.

\begin{figure}[h]
    \centering
    \includegraphics[scale=0.5]{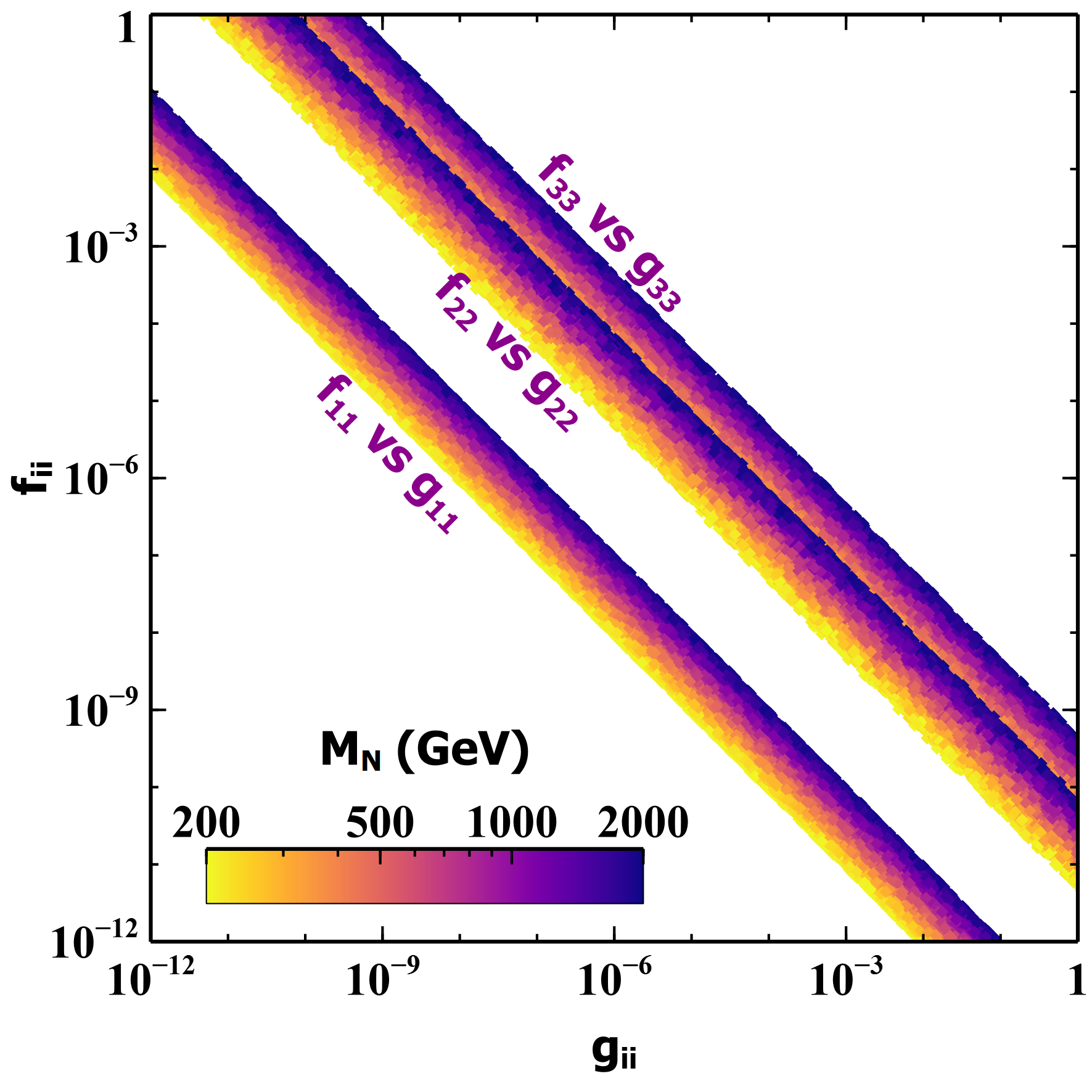}
    \caption{The range of diagonal elements of  $\textbf{f}$ and $\textbf{g}$ matrices which satisfy the neutrino oscillation data. Here we have used $\langle \eta\rangle =u = 1$ GeV.}
    \label{fig:coupling}
\end{figure}

\section{Contribution to $\Delta N_{\rm eff}$}\label{sec::neff}

As mentioned earlier, the insistence on the Dirac nature of neutrinos dictates that the newly introduced right-chiral fermions $\nu_R$s have a mass similar to SM neutrinos. The existence of these additional ultra-light species in the early Universe can significantly contribute to the total radiation energy density. Consequently, they affect the effective relativistic degrees of freedom denoted as $N_{\rm eff}$ which is expressed as 
\begin{equation}
N_{\rm eff} \equiv \frac{\rho_{\rm rad} - \rho_{\gamma}}{\rho_{\nu_L}}.
\label{eq:Neff}
\end{equation}
Here, $\rho_{\rm rad}$ signifies the total energy density of the thermal plasma, while $\rho_{\gamma}$ and $\rho_{\nu_L}$ represent the energy density of photons and a single active neutrino species, respectively. In the absence of any novel light degrees of freedom, the Standard Model precisely predicts $N_{\rm eff}$ and is commonly quoted as $3.045$~\cite{Mangano:2005cc,Grohs:2015tfy,deSalas:2016ztq,Cielo:2023bqp, Akita:2020szl, Froustey:2020mcq, Bennett:2020zkv}.
In Dirac neutrino mass models, because of the presence of $\nu_R$s, the additional contribution to $N_{\rm eff}$, in the total radiation energy density can be written as,
	\begin{eqnarray}
	\Delta{N_{\rm eff}}&=& N_{\nu_R}\times \frac{\rho_{\nu_R}}{\rho_{\nu_L}} \Bigg|_{\rm T=T_{\rm CMB}},
	\end{eqnarray}
	where $N_{\nu_R}$ is the number of generations of $\nu_R$, and $\rho_{\nu_R}$ is the energy density of the single generation of $\nu_R$ where we assume that all three $\nu_R$s behave identically and hence contribute equally to the energy density. 

 In our setup, $\nu_R$ establishes a connection with the SM bath solely through the coupling term $g_{ij} \overline{N}_{L_i} \eta \nu_{R_j}$. Consequently, the production of $\nu_R$ in the early Universe is contingent solely on the Yukawa coupling $g$. The strength of $g$ dictates the possibility of both thermal and non-thermal production of $\nu_R$. In Fig.~\ref{fig:coupling}, we illustrate the relative coupling strength of the diagonal elements of $\textbf{f}$ and $\textbf{g}$ that satisfies the neutrino oscillation data, with $M_N$ varying in the range $[200,2000]$ GeV, depicted in the color code. 

\begin{figure}[h]
		\centering
		\includegraphics[width=0.2\textwidth]{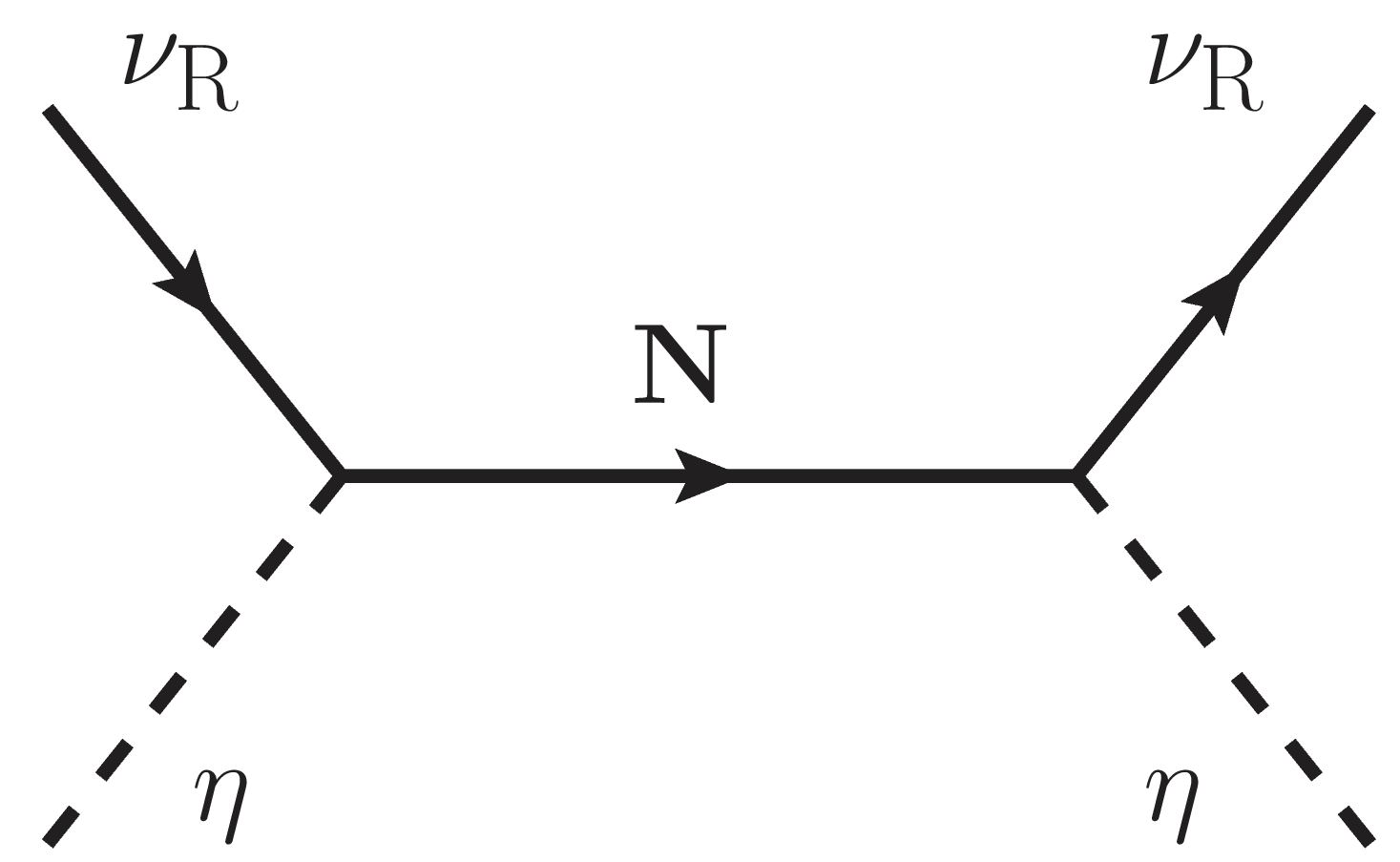}
        \includegraphics[width=0.2\textwidth]{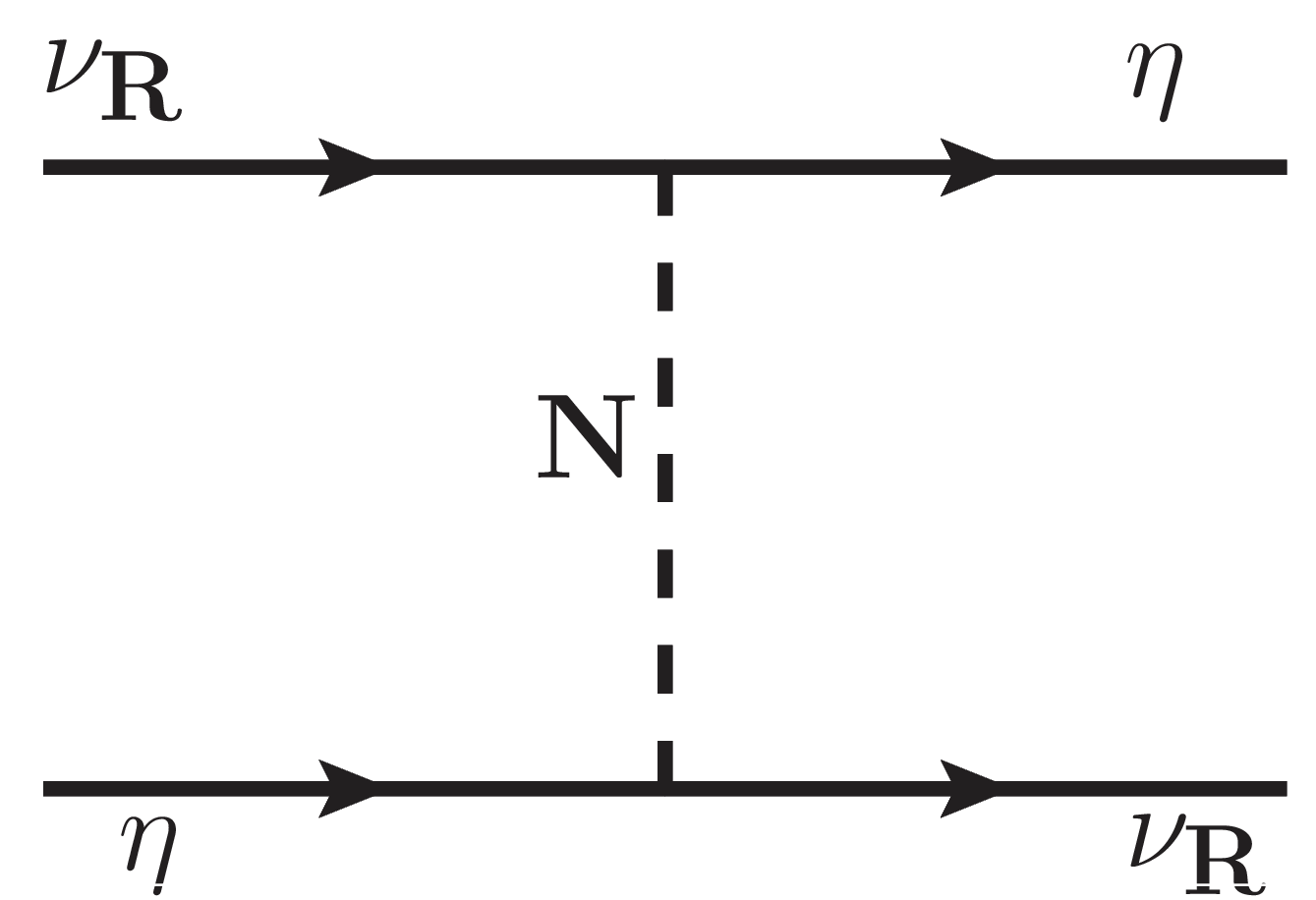}
        \includegraphics[width=0.2\textwidth]{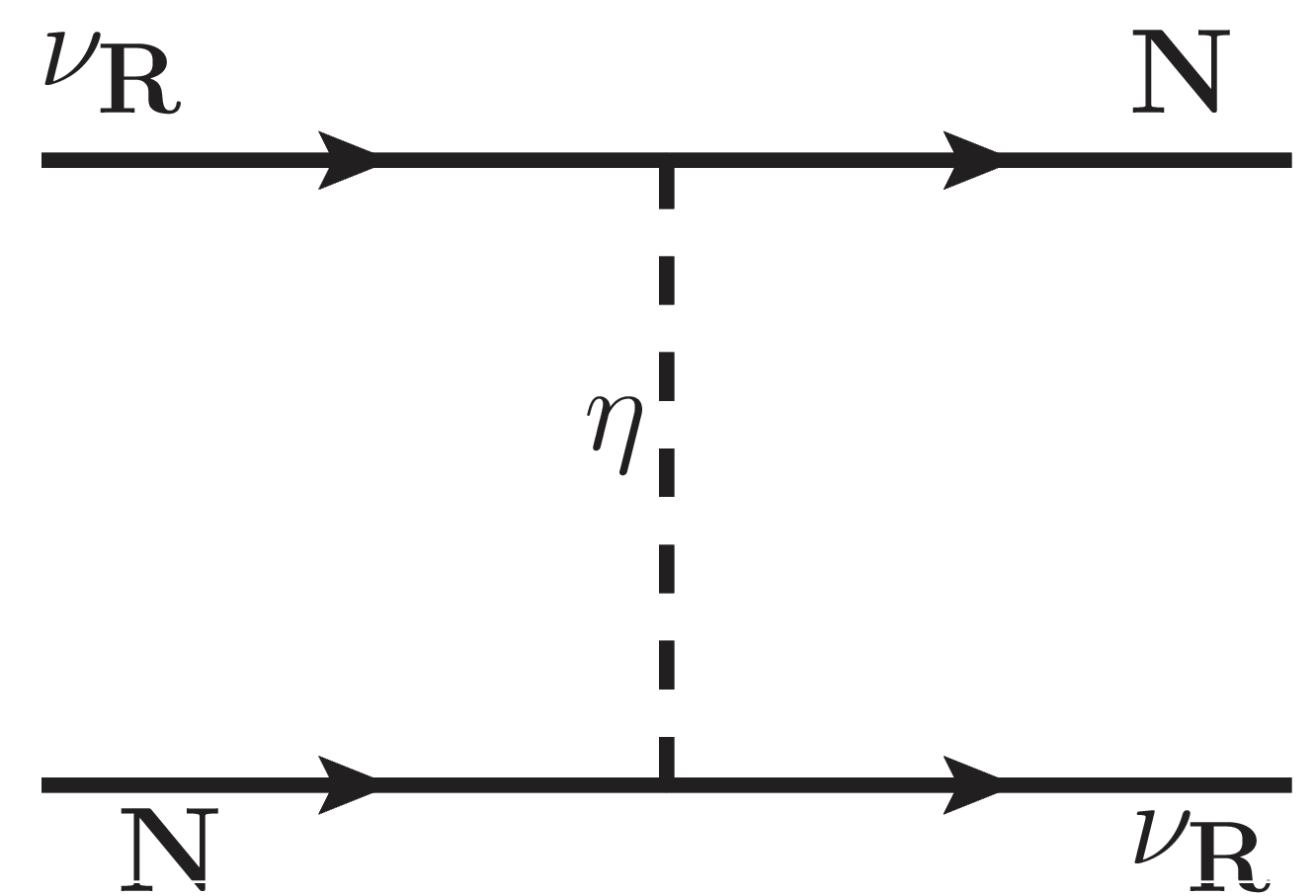}
		\caption{Elastic scattering of $\nu_R$ with $\eta$ and $N$, to keep $\nu_R$ in thermal equilibrium. }
		\label{fig:nrelastic}
	\end{figure}

\begin{figure}[h]
		\centering
		\includegraphics[scale=0.45]{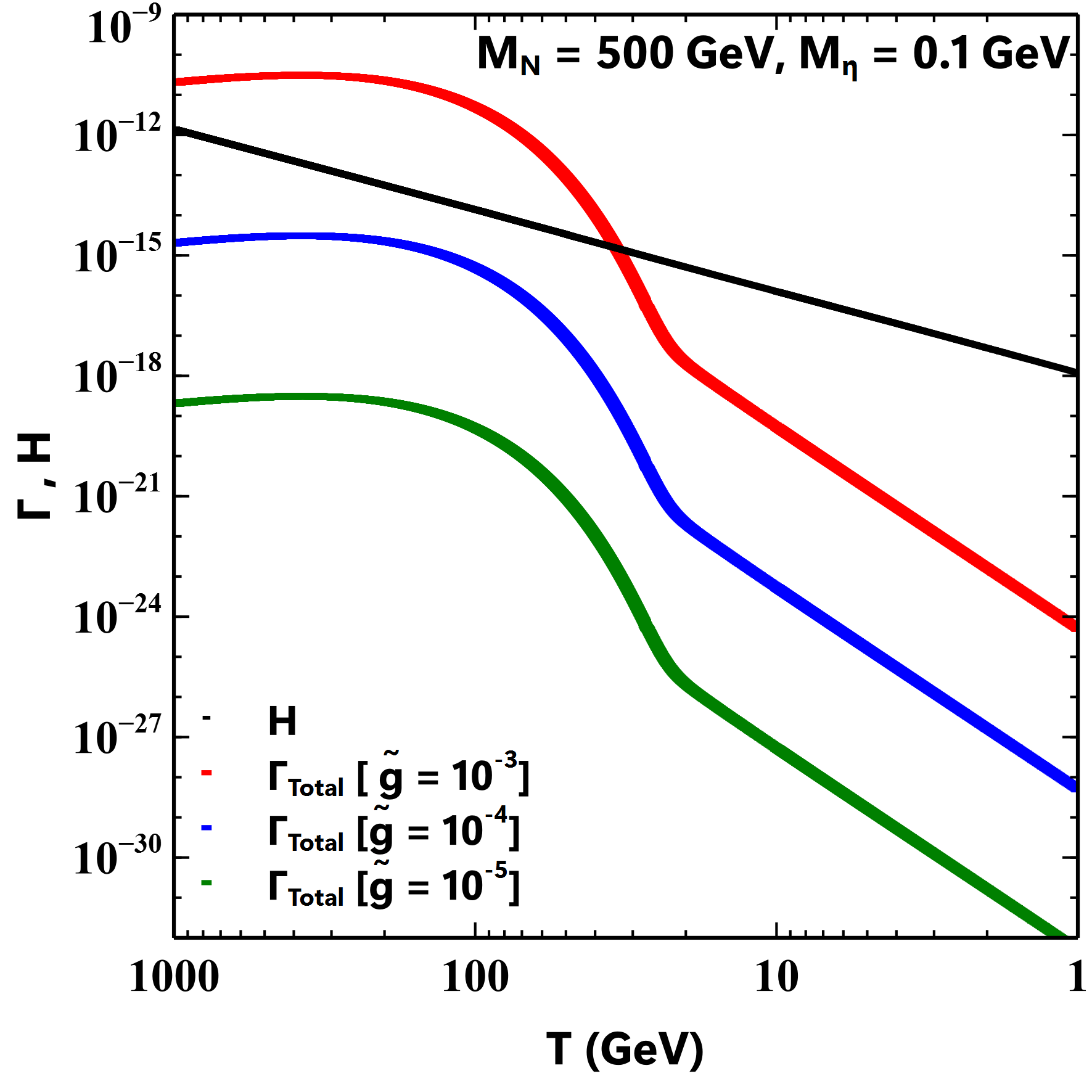}
		\caption{${\Gamma_{\rm Total}=(\Gamma_{\eta \nu_R \leftrightarrow \eta \nu_R}+\Gamma_{N \nu_R \leftrightarrow N \nu_R})}$ and $ {H}$ are plotted as a function of $T$. We have taken three values of the  coupling $\Tilde{g}$ = $10^{-3}, 10^{-4}, 10^{-5}$, $M_N = 500$ GeV and $M_\eta = 0.1$ GeV. }
		\label{fig:etanr1}
	\end{figure}

Fig. \ref{fig:nrelastic} delineates the processes involved in maintaining $\nu_R$ in thermal equilibrium. By comparing the interaction rate of these pertinent processes against the Hubble expansion rate, we estimate the lower limit on the coupling $g$ necessary to keep $\nu_R$ in equilibrium with the thermal bath. This has been demonstrated in Fig.~\ref{fig:etanr1} for a set of benchmark values of the coupling parameter $\Tilde{g}={\sum_i g_{ii}}/{3}=10^{-3}, 10^{-4}, 10^{-5}$ by keeping the other parameters $M_{N}$ and $M_\eta$ fixed at $500$ GeV and $0.1$ GeV respectively. 

As the significance of this contribution can differ based on whether the $\nu_R$s existed in the thermal bath or were generated non-thermally~\cite{Luo:2020sho, Luo:2020fdt, Biswas:2022fga,Biswas:2021kio,Biswas:2022vkq,Borah:2023dhk}, based on this criterion, we categorize our analysis into two scenarios: (i){ [Case-1]}: $\Tilde{g}\geq 10^{-3}$ (Thermal Production) and 
	(ii) { [Case-2]}: $\Tilde{g}<10^{-3}$ (Non-thermal Production).
 
 \vspace{0.3cm}
\noindent\underline{\textbf{\large Case-1}} [$10^{-3}\leq\Tilde{g}<\sqrt{4\pi}$]:\\

In this scenario, the substantial interaction rate ensures that $\nu_R$ remains in thermal equilibrium along with $N$ and $\eta$. The primary production mechanism for $\nu_R$ arises from the annihilation of $\eta$ and $N$. As long as the interaction rate of elastic scattering processes (depicted in Fig. \ref{fig:nrelastic}) exceeds the Hubble expansion rate, $\nu_R$ stays in equilibrium with the thermal bath. Once this interaction rate drops below the Hubble parameter, $H(T)$, the $\nu_R$ species decouple from the thermal bath and evolve independently. The energy density of $\nu_R$ at the time of decoupling, determined by their decoupling temperature, contributes to $\Delta N_{\rm eff}$ and is given by:

\begin{equation}\label{eq::deln1}
\Delta N_{\rm eff} = N_{\nu_R}\times\left(\frac{T_{\nu_R}}{T_{\nu_L}}\right)^4 = N_{\nu_R}\left(\frac{g_{*s}(T^{\rm dec}_{\nu_L})}{g_{*s}(T^{\rm dec}_{\nu_R})}\right)^{4/3},
\end{equation}

where, $g_{*s}(T_{\alpha}^{\rm dec})$ represents the relativistic entropy degrees of freedom at the decoupling temperature $T_{\alpha}^{\rm dec}$ for the species $\alpha$ (where $\alpha = \nu_L, \nu_R$).

\begin{figure}[h]
		\centering
  \includegraphics[scale=0.45]{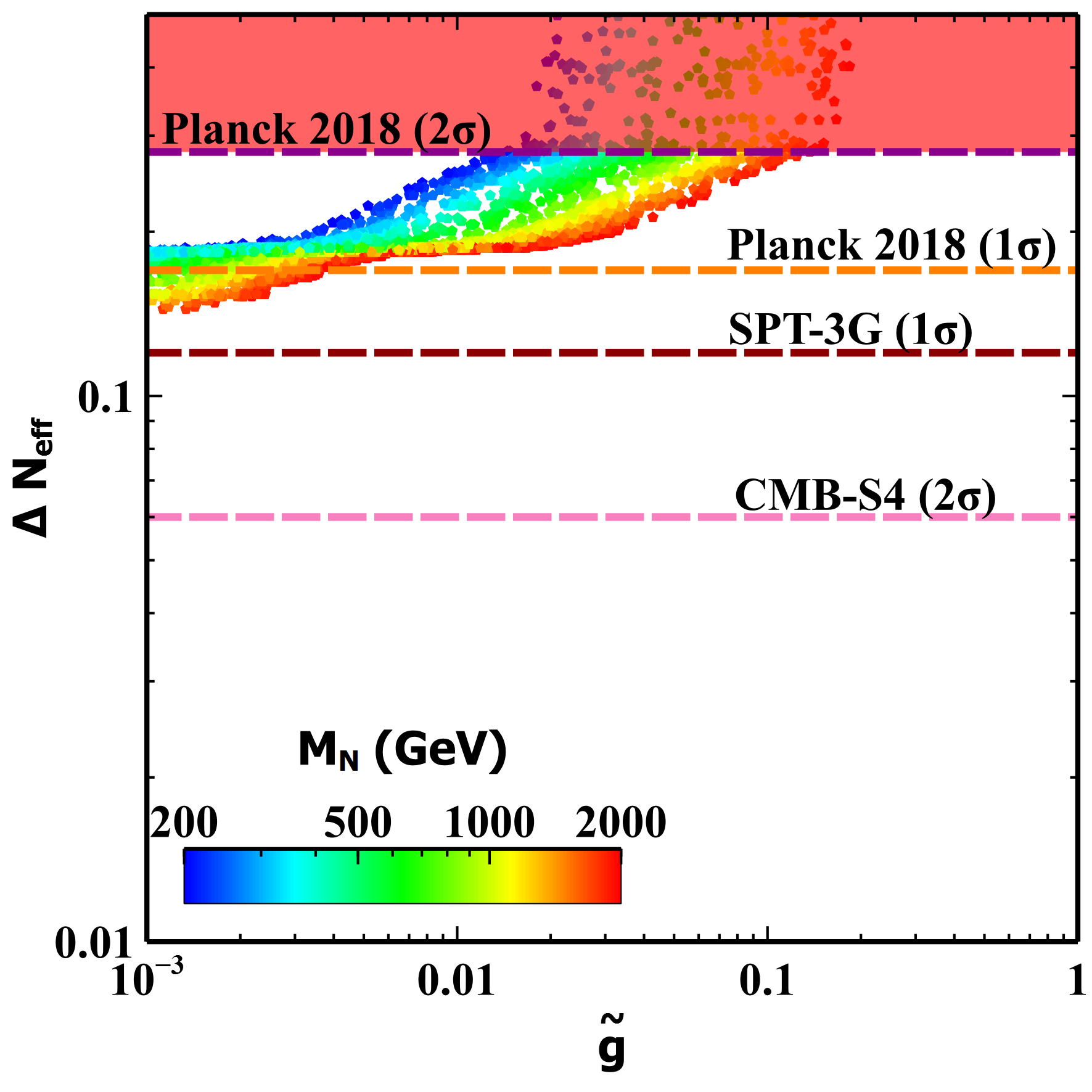}
		\caption{The contribution to $\Delta N_{\rm eff}$ is shown from the thermal production of $\nu_R$ with respect to the coupling $\Tilde{g}$. The colour code represents the value of $M_N$.}
		\label{fig::c1deln}
	\end{figure}

Fig. \ref{fig::c1deln} illustrates the contribution of $\Delta N_{\rm eff}$ from the $\nu_R$s that were once in equilibrium with the SM plasma, as a function of the coupling $\Tilde{g}$. This has been calculated by using Eq.~\ref{eq::deln1}, after evaluating the decoupling temperature of $\nu_R$ from the thermal plasma. The plot demonstrates an increase in the $\Delta N_{\rm eff}$ contribution with a higher coupling strength $\Tilde{g}$. This is due to the fact that a larger $\Tilde{g}$ results in a higher interaction rate, maintaining $\nu_R$ in thermal equilibrium for an extended period, causing a relatively late freeze-out of $\nu_R$ and consequently contributing more to $\Delta N_{\rm eff}$. The color-coded representation of $M_N$ indicates that as the mass of $N$ increases, the interaction rate decreases, and freeze-out occurs earlier. Therefore, a heavier $M_N$ leads to a lower contribution to $\Delta N_{\rm eff}$. The red shaded region in the plot is already excluded by the Planck-2018 data at $2\sigma$ C.L.~\cite{Planck:2018vyg}.

We also observe that once three of the $\nu_R$s were produced in the thermal bath, $\Delta N_{\rm eff}$ would always have a minimum contribution of $0.14$, well above the future sensitivity of CMB-S4 and SPT-3G~\cite{Abazajian:2019eic,SPT-3G:2019sok}. As CMB-S4 or SPT-3G can probe $\Delta N_{\rm eff}$ down to 0.06 at $2\sigma$ CL, they have the potential to validate or falsify this scenario.

\vspace{0.3cm}
 
 \noindent \textbf{\underline{\large Case-2} [$10^{-12}<\Tilde{g}<10^{-3}$]:}\\
 
 \begin{figure}[h]
		\centering
		\includegraphics[width=0.4\textwidth]{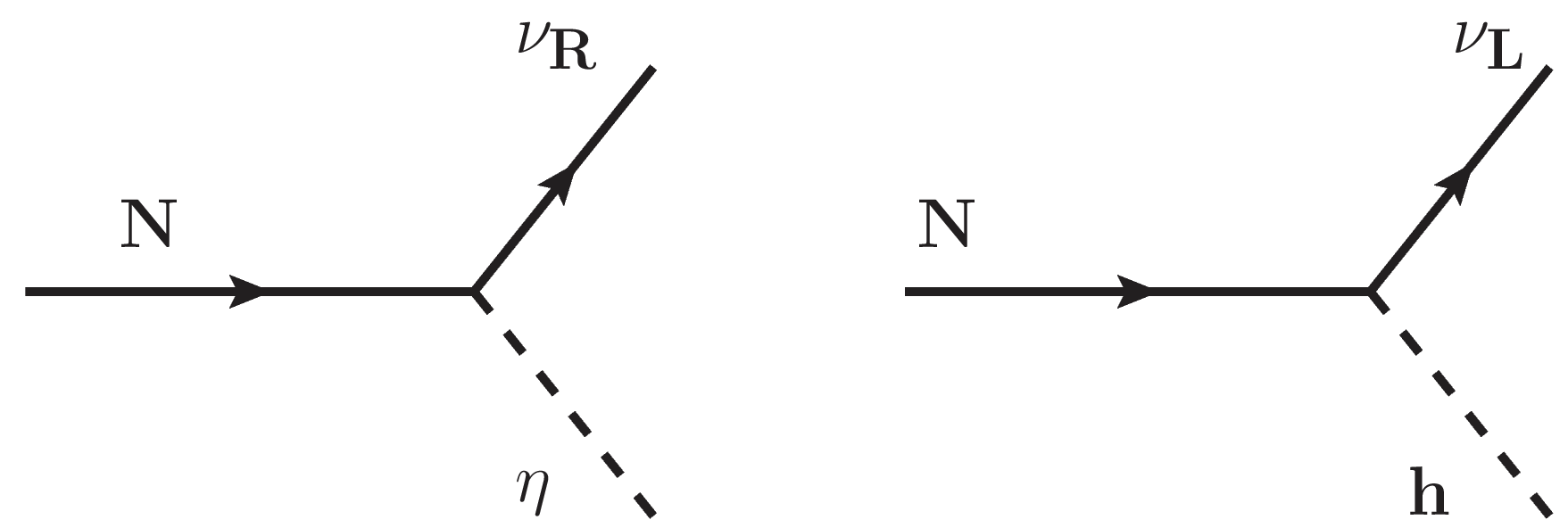}
		\caption{The Feynmann diagram for two possible decay processes of N.}
		\label{fig:Ndecay}
	\end{figure}
 
In this scenario, the interaction strength of $\nu_R$ is notably weak, making thermal production of $\nu_R$ infeasible. However, through $N_k$ decays, it becomes possible to generate sufficient $\nu_R$ energy density to meet the current sensitivity of $\Delta N_{\rm eff}$. Since $\Tilde{g}$ is small in this case, the coupling $\Tilde{f}_k=f_{k1}+f_{k2}+f_{k3}$ can be large while satisfying the constraints from neutrino oscillation data as demonstrated in Fig.~\ref{fig:coupling}. Thus $\Tilde{f}_k$ coupling plays a crucial role in maintaining the thermal equilibrium of $N_k$ with the SM plasma. To ensure the thermal equilibrium of $N_k$ with the bath particles ($\Phi$ and $\nu_L$), we keep $\Tilde{f}$ in the order $\mathcal{O}(10^{-3})$ or larger for $M_N \sim \mathcal{O}(10^2-10^3)$ GeV.
Thus we consider the scenario in which $N_k$ was in equilibrium in the early Universe, and from its decay, we will evaluate the abundance of $\nu_R$. The two decay channels for $N_k$ are shown in the Fig. \ref{fig:Ndecay}. To track the evolution of $\nu_R$ and $N_k$, the relevant Boltzmann equations can be written as:
	\begin{eqnarray}
	\frac{dY_{N_{k}}}{dx}&=&\frac{\beta s}{Hx} \left[- \langle\sigma v\rangle_{N_{k} N_{k}\rightarrow X \bar{X}}(Y^2_{N_{k}}-(Y^{\rm eq}_{N_{k}})^2)\right.\nonumber\\
  &-& \left.\frac{\Gamma_2}{s}\frac{K_1(x)}{K_2(x)}Y_{N_{k}}- \frac{\Gamma_1}{s}\frac{K_1(x)}{K_2(x)}(Y_N-Y_{N_{k}}^{eq})\right]
  \label{eq::BEforN}
	\end{eqnarray}
	\begin{equation}
	\frac{d\Tilde{Y}_{\nu_{Rk}}}{dx}=\frac{\beta}{Hs^{1/3}x}\langle E\Gamma_2\rangle Y_{N_k},
    \label{eq::BEfornr}
	\end{equation}
	
	where the dimensionless parameters $Y_{N_{k}}={n_{N_{k}}}/{s} ~{\rm and}~\Tilde{Y}_{\nu_{Rk}}={\rho_{\nu_{Rk}}}/{s^{4/3}}$. $\Tilde{Y}_{\nu_{Rk}}$ is the $\nu_R$ abundance produced from $N_k$ decay. In the above equation $\Gamma_{1}$, $\Gamma_{2}$ and $\langle E\Gamma_2\rangle$ can be expressed as
	\begin{eqnarray}
 \Gamma_1({N_{k}\rightarrow h \nu_{L}})&=&\frac{\sum_{j} (f_{kj})^2}{8\pi} M_{N_{k}} \left(1-\frac{M_{h}^2}{M_{N_{k}}^2}\right)^2\label{eq::gamma2}\\
	\Gamma_2({N_{k}\rightarrow\eta \nu_R})&=&\frac{g_{kk}^2}{8\pi} M_{N_{k}} \left(1-\frac{M_{\eta}^2}{M_{N_{k}}^2}\right)^2\\
	\langle E\Gamma_2({N_{k}\rightarrow\eta \nu_R})\rangle&=&\frac{{g_{kk}^2}}{16\pi} M_{N_{k}}^2\left(1-\frac{M_{\eta}^2}{M_{N_{k}}^2}\right)^2\label{eq::egamma2}
	\end{eqnarray}	
 
Here, we have used index $k$ for the three generations of N and index $j$ in $f_{kj}$ is for three generations of $\nu_L$. In Eq.~\ref{eq::BEforN}, the $\langle\sigma v\rangle_{N_{k} N_{k}\rightarrow X \bar{X}}$ is the total thermal averaged cross section for annihilation of $N_{k}$ into all the particles that are kinematically accessible. The second term is for decay of $N_{k}$ to $\eta$ and $\nu_R$ while the third term corresponds to the decay and inverse decay of $N_k$ to $h$ and $\nu_L$. The inverse decay is not included in the second term as $\nu_R$ is never in thermal equilibrium and since its number density is extremely small as compared to the bath particles, it can be ignored.  In this scenario, as $\Tilde{f}_k$ is larger than $g_{kk}$ and is crucial for keeping the $N_k$ in equilibrium, $N_k$ always dominantly decays to $h$ and $\nu_L$. For $g_{kk}\leq10^{-5}$ and $\Tilde{f}_k\geq10^{-3}$, even though the branching ratio for $N_k\rightarrow \eta \nu_R$ is always suppressed ({\it i.e.} ${\rm Br}(N_k\to\eta \nu_R)\ll 1\%$), it is still possible that significant amount of $\nu_R$ can be produced untill the inverse decay $\nu_L \phi\to N_k$ is effective maintaining the $N_k$ number density.  Once $x(={M_{N_{k}}}/T)$ goes beyond $1$, because of the Boltzmann suppression, this inverse decay will no longer be effective 
and hence the $\nu_R$ production from $N_k$ decay will be suppressed.
 After solving the Boltzmann Eqs.~\ref{eq::BEforN} and \ref{eq::BEfornr} for three generations of N, $\Delta N_{\rm eff}$ produced by $\rho_{\nu_{Rk}}$ can be calculated as:
	\begin{eqnarray}\label{eq::deln2}
	\Delta N_{\rm eff}&=& 2\left(\frac{\sum_{k=1}^{3}\rho_{\nu_{Rk}}}{\rho_{\nu_L}}\right)_{T_D(\nu_L)\nonumber}\\
	&=&2\left(\frac{s^{4/3}\sum_{k=1}^{3}\Tilde{Y}_{\nu_{Rk}}}{\rho_{\nu_L}}\right)_{T_D(\nu_L)},
	\end{eqnarray}
  where $T_D(\nu_L)$ is the standard neutrino decoupling temperature. The factor 2 is for $\nu_R$ and anti-$\nu_R$.
  
\begin{figure}[h]
		\centering
  \includegraphics[scale=0.40]{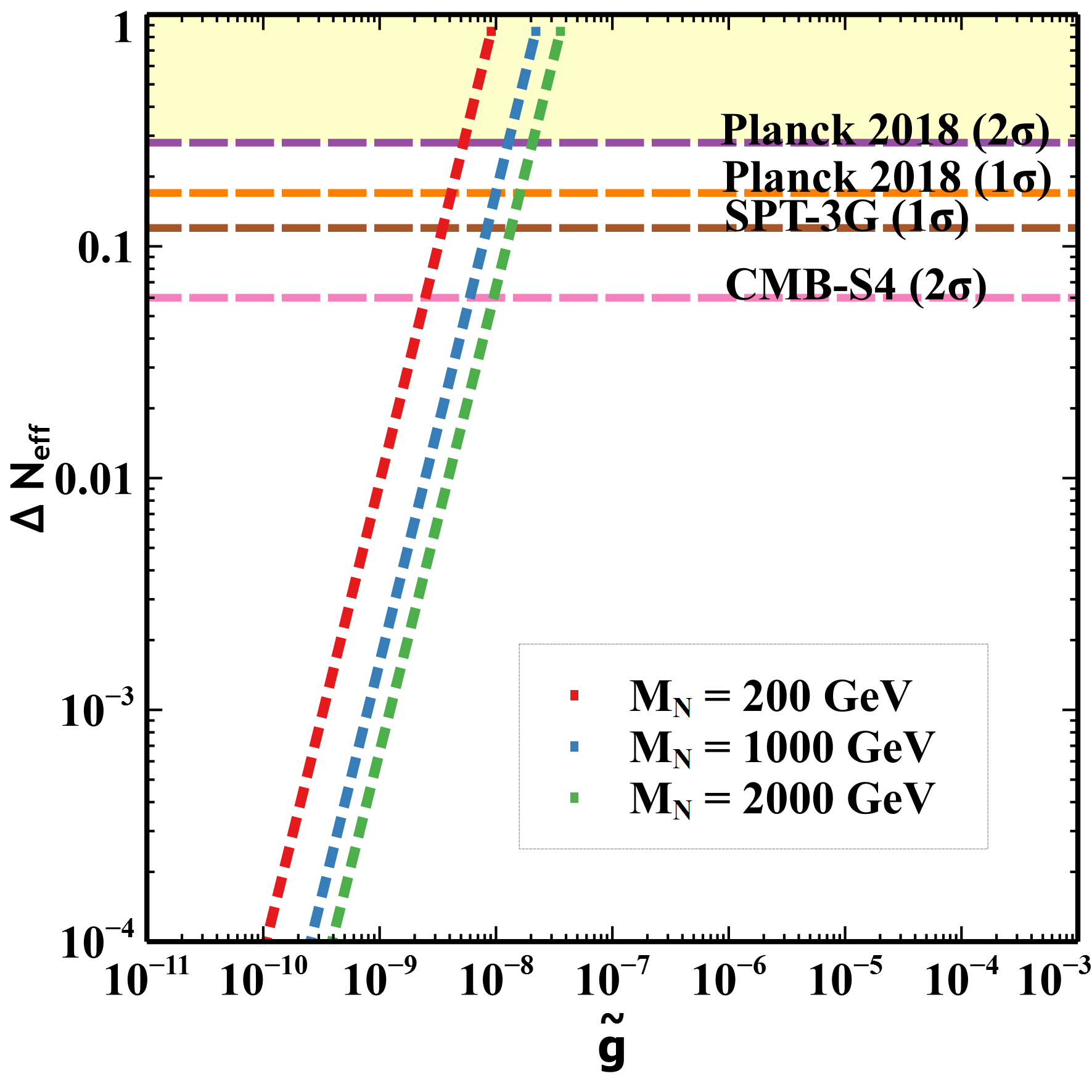}
		\caption{The contribution to $\Delta N_{\rm eff}$ is shown from the Freeze-in production of $\nu_R$ with respect to the coupling $g$. Three benchmark values for the mass of N are taken. $M_N$ ($=M_{N_1}$$=M_{N_2}$$=M_{N_3}$) = 200 GeV, 1000 GeV and 2000 GeV correspond to red, blue and green respectively. $M_\eta=0.1 GeV.$ }
		\label{fig:c2deln}
	\end{figure}
 
In Fig. \ref{fig:c2deln}, the contribution to $\Delta N_{\rm eff}$ from non-thermal production of $\nu_R$ is illustrated. It is evident that an increase in the coupling strength $\Tilde{g}$ leads to a higher decay width, resulting in enhanced $\nu_R$ production. According to Eq. \ref{eq::egamma2}, the decay width also rises with \(M_N\). One might anticipate that, with increasing \(M_N\), more $\nu_R$ would be produced, consequently yielding a larger $\Delta N_{\rm eff}$. However, the plot reveals a contrary trend where $\Delta N_{\rm eff}$ decreases as \(M_N\) increases.
This seemingly counterintuitive behavior can be rationalized by considering the production timescale of $\nu_R$. For larger \(M_N\), $\nu_R$ is produced earlier in the Universe compared to scenarios with lighter \(M_N\). The energy density of early-produced $\nu_R$ experiences more significant dilution due to redshift compared to later-produced $\nu_R$ energy density. The yellow region in the plot is excluded by Planck-2018 data at \(2\sigma\) C.L., while future experiments such as SPT-3G and CMB-S4~~\cite{Abazajian:2019eic,SPT-3G:2019sok} are poised to probe portions of the $\Delta N_{\rm eff}$ parameter space depicted in the plot.

\section{Self-Interacting DM} \label{sec::SIDM}

Self-interacting dark matter (SIDM) emerges as a solution to address small-scale anomalies encountered in the $\Lambda$CDM model. These anomalies, including the 'core-cusp problem' concerning the density profiles of dark matter halos in galaxies \cite{de2010core}, the 'too big to fail' problem associated with the absence of the most luminous satellite galaxies in the most massive sub-halos \cite{Bullock:2017xww,boylan2011too,boylan2012milky,Tulin:2017ara}, and the 'missing-satellite problem' involving the over-prediction of small satellite galaxies in simulations \cite{Bullock:2017xww,Moore:1999nt,Klypin:1999uc}, reveal discrepancies between the predictions of $\Lambda$CDM and observations on smaller scales.

Unlike the standard $\Lambda$CDM model, which considers dark matter as collision-less, SIDM allows dark matter particles to interact with each other through self-scattering, extending beyond gravitational interactions. This interaction in SIDM involves elastic scattering through a t-channel, mediated by either a gauge boson or a scalar particle. The normalized cross-section is constrained by observations \cite{Markevitch:2003at,Clowe:2003tk,Randall:2008ppe,Harvey:2015hha} and is approximately within the range of $\sigma/m_{\rm DM} \sim (0.1-1){\rm cm^2/g}$ for clusters (with velocities around $1000$ km/s), $(0.1-10){\rm cm^2/g}$ for galaxies (with velocities around $200$ km/s), and $(0.1-100)~{\rm cm^2/g}$ for dwarf galaxies (with velocities around $10$ km/s).  

\begin{figure}[h]
    \centering
    \includegraphics[scale=1.4]{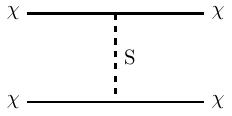}
    \caption{Feynman Diagram depicting the dark matter self-interaction.}
    \label{fig:sidm}
\end{figure}

In this paper, we explore the possibility of realization of a fermionic SIDM $\chi$ mediated by a scalar particle $S$ both of which are charged under the 
$Z_4$ symmetry. The stability of the dark matter is further guaranteed by the remnant $Z_2$ symmetry, under which $\chi$ is odd while all the other particles are even. 
The diagram illustrating the self-interaction process is shown in Fig. \ref{fig:sidm}. In this scenario, the non-relativistic self-interaction of dark matter (DM) is effectively described by a Yukawa-type potential: $V(r)= \frac{y^2_{\chi}}{4\pi r}e^{-M_{S}r}$. For details on the self-interaction cross-section, please see Appendix~\ref{app::SIDM cross-sections}. 

\subsection{DM Relic Density}\label{sec::Relic}
\begin{figure}[h]
    \centering
    \includegraphics[scale=1.4]{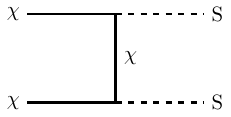}
    \caption{Feynman diagram for the dark matter annihilation.}
    \label{fig:relic}
\end{figure}

In the outlined model, the scalar particle $S$ serves as the mediator for self interactions of dark matter. It also establishes a portal between DM and visible sector through its coupling with the SM Higgs and $\eta$. The scalar couplings $\lambda_{hs}$ and $\lambda_{\eta s}$ establish thermal connections, bringing the $S$ particle into equilibrium with the SM bath. This thermal equilibrium facilitates the freeze-out mechanism, crucial for achieving the required relic density of DM. In order to ensure adequate self-interaction among DM particles, a substantial coupling and a relatively smaller mediator mass is necessary. It  consequently gives rise to the dominant channel governing the DM freeze-out process, {\it i.e.} the annihilation process $\chi \chi \rightarrow S S$ which is depicted in Fig. \ref{fig:relic}. This process results in significant DM annihilation rates, often leading to a lower-than-desired relic abundance in the low DM mass range~\cite{Borah:2022ask}. While achieving a pure thermal relic poses challenges,  recent studies have delved into a hybrid approach that incorporates both thermal and non-thermal contributions. This approach, potentially leading to the correct relic density for SIDM, introduces new degrees of freedom, rendering the model non-minimal\cite{Borah:2022ask,Borah:2021yek, Borah:2021pet, Borah:2021rbx, Borah:2021qmi}. Here, we aim to achieve the correct relic density in the most minimal setup without adding any new degrees of freedom.

We note that, if $\lambda_{hs}$ and $\lambda_{\eta s}$ is greater than $\mathcal{O}(10^{-6})$, then it ensures that $S$ stays in equilibrium and consequently ensures the thermal equilibrium of DM with the SM plasma. 
The relevant Boltzmann equations to track the comoving number density of DM and  $S$ can be written as:
\begin{equation}\label{eqn::Boltzmann}
    \begin{split}
        \frac{dY_{\chi}}{dx} = & -\frac{s(M_{\chi})}{H(M_{\chi})} \frac{\langle \sigma v \rangle_{\chi \chi \rightarrow S S}}{x^2}\left(Y_{\chi}^2-Y_{S}^2\right) \\
        \frac{dY_{S}}{dx} = & ~ \frac{s(M_{\chi})}{H(M_{\chi})} \frac{\langle \sigma v \rangle_{\chi \chi \rightarrow S S}}{x^2}\left(Y_{\chi}^2-Y_{S}^2\right)\\
        & - \frac{s(M_{\chi})}{H(M_{\chi})} \frac{\langle \sigma v \rangle_{S S \rightarrow \eta \eta}}{x^2}\left(Y_{S}^2-(Y^{eq}_{S})^2\right)\,,
    \end{split}
\end{equation}
where $s(M_\chi)$ and $H(M_\chi)$ are the entropy density and Hubble rate respectively and $x$ is the dimension less parameter defined as $x=M_{\chi}/T$. 
\noindent The thermally averaged cross-section for the DM annihilation to the scalar mediator is given by 
\begin{equation}
    \langle \sigma v \rangle=\frac{3}{4}\frac{y_{\chi}^2}{16 \pi M_\chi^2}v^2\sqrt{1-\frac{M^2_S}{M_\chi^2}}
\end{equation}

\begin{figure}[h]
    \centering
    \includegraphics[scale=0.45]{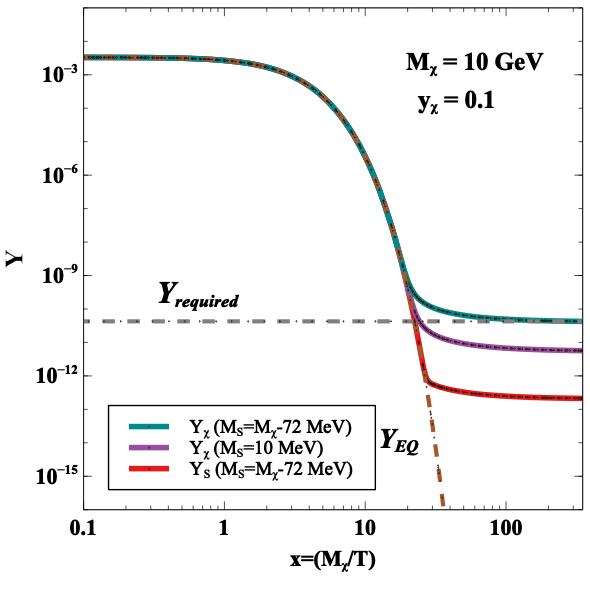}
    \caption{Evolution of co-moving number densities of DM and $S$.}
    \label{fig::relic_plot}
\end{figure}

In Fig.~\ref{fig::relic_plot}, we illustrate the evolution of the comoving number densities of $\chi$ and $S$. We set the DM mass as $M_\chi=10$ GeV and the DM self-interaction coupling as $y_{\chi}=0.1$. The comoving number density of $\chi$ with $M_S=9.928$ GeV is represented by the green solid line, while the one with $M_S=0.01$ GeV is depicted by the purple solid line. Evidently, by adjusting the mass of $S$, it is possible to achieve the correct relic density of $\chi$. This mass tuning of the scalar $S$ is facilitated by a phase transition~\cite{Elor:2021swj,Cohen:2008nb,Hashino:2021dvx,Borah:2023sal}. Before the phase transition, the mass of $S$ is such that $M^i_S\lesssim M_{\rm DM}(=M_\chi)$, ensuring that the DM annihilation rate to $S$ is phase-space suppressed. This reduction in the $\chi\chi \to SS$ annihilation cross-section leads to the correct relic density of $\chi$. Subsequent to the phase transition, $S$ becomes light with $M_{S} \sim \mathcal{O}(10)$ MeV, which is necessary to explain the small scale problems (sub-Galactic scale) through self-interaction of DM. This scenario is feasible if the mediator $S$ is coupled to another scalar $\xi$, inducing a first-order phase transition. Through a coupling term like $\mu \xi S^\dagger S$, below the nucleation temperature of the FOPT, the physical mass of the mediator can undergo a change to $(M^f_S)^2 =(M^i_S)^2 -\mu {v_\xi}$, where $v_\xi$ represents the vacuum expectation value acquired by $\xi$. Careful fine-tuning between the two terms $(M^i_S)^2$ and $\mu {v_\xi}$ allows achieving a final mediator mass suitable enough to achieve the required self-interaction. 



As discussed earlier, the scalar $S$ does not mix with the SM Higgs $h$. Consequently, $S$ does not decay into SM particles.
However, $S$ can annihilate to $\eta$ particle if $M_\eta < M_S^i$, resulting in negligible relic of $S$.
In Fig.~\ref{fig::relic_plot}, the comoving number density of $S$ is shown by the red solid line for the process $SS \rightarrow \eta \eta$ by choosing a typical value of $\lambda_{\eta S}=0.6$. The $\eta$ particle mixes with the SM Higgs $h$ and decay to the SM particles well before the Big Bang Nucleosynthesis (BBN). Further discussion on lifetime of $\eta$ particle is given in Section \ref{sec::somecon}. Consequently, the total dark matter relic density in the Universe is solely comprised of the $\chi$ abundance.

\subsection{Direct Detection}
\begin{figure}[h]
    \centering
    \includegraphics[scale=01]{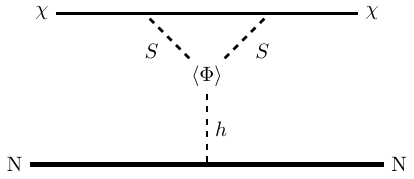}
    \caption{Feynman diagram for dark matter interaction with the nucleon through one-loop process.}
    \label{fig::detection}
\end{figure}

The possibility of spin-independent DM nucleon elastic scattering allows for the detection of DM in terrestrial laboratories. As $S$ does not acquire a vev, it does not mix with SM Higgs and hence the tree-level DM-nucleon scattering through the $S-H$ mixing portal is not possible as compared to \cite{Borah:2021rbx, Borah:2022ask}. Thus, in our case, the simplest diagram for direct detection is at the one-loop level, as depicted in Fig.~\ref{fig::detection}.

\begin{figure}[h]
    \centering
    \includegraphics[scale=0.5]{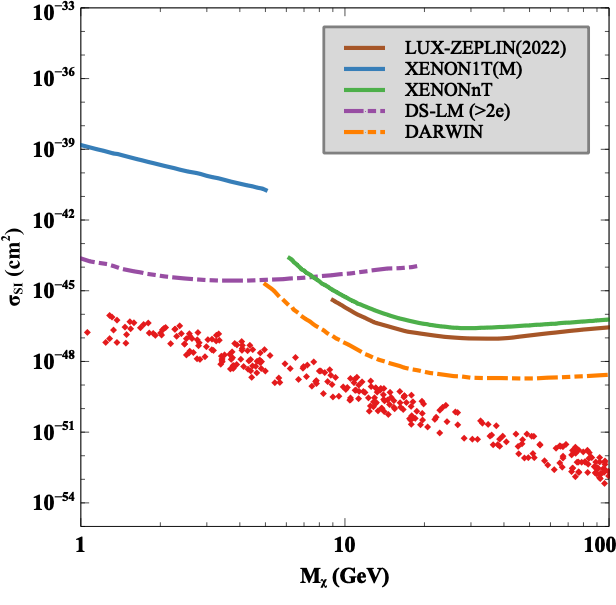}
    \caption{Spin-Independent DM-nucleon scattering cross-section as a function of DM mass, $M_\chi$.}
    \label{fig::DD_bound}
\end{figure}

The Higgs exchange diagram induces an effective scalar interaction term between the dark matter $\chi$ and the quark $q$ of the form $C_q^{tri}\,\overline{q}q\overline{\chi}\chi$, with 
\begin{equation}
    C_q^{tri}=\frac{y_{\chi}^2}{16\pi^2 M_h^2 M_\chi}\left[\frac{\lambda_{hs}}{2}L\left(\frac{M_\chi^2}{M_S^2} \right)\right]m_q
\end{equation}
Here the loop function $L(x)$ is given by 
\begin{equation}\label{app::loop}
        L(x)  = (1+x^{-1})ln(1+x)-1\,,
\end{equation}
the details of which is given in the Appendix~\ref{app::loop}. Then, the spin-independent scattering cross-section of $\chi$ off the nucleons can be expressed as

\begin{equation}\label{eqn::cross-section}
    \sigma_{\rm SI}=\frac{1}{\pi} \frac{M_\chi^2 m_n^2}{(M_\chi +m_n)^2} m_n^2 f^2 \left(\frac{C_q^{tri}}{m_q}\right)^2
\end{equation}

where we have considered $f$($=f_p$=$f_n$)= 0.308 \cite{Hoferichter:2017olk}, and $m=m_p,m_n$ is the nucleon mass. Fig.~\ref{fig::DD_bound} shows the evaluated cross-section (as shown in scattered points) against various experimental data.

In Fig.~\ref{fig::DD_bound}, we showcase the $\sigma_{\rm SI}$ calculated from Eq.~\ref{eqn::cross-section}, for the points giving rise to required self-interaction, as a function of DM mass by the red colored points. We also present the existing constraints from LUX-ZEPLIN (LZ) experiment \cite{LZ:2022lsv}, the XENON1T (Migdal) \cite{XENON:2019zpr} and XENONnT \cite{XENON:2023cxc} and future sensitivities of DARWIN \cite{DARWIN:2016hyl} and DS-LM \cite{GlobalArgonDarkMatter:2022ppc} direct search experiments by different colored solid lines and dot-dashed lines respectively. For the scan, keeping $\lambda_{hs}$ fixed at $\lambda_{hs}=10^{-2}$, we vary $y_{\chi}$ in a range $\{0.5,1\}$ and $M_S$ is also varied in a range $\{10,100\}$ MeV.   
Clearly the SIDM parameter space remains safe from DM direct search constraints and lies beyond the reach of future sensitivities. Hence this again emphasizes the importance of observable $\Delta N_{\rm eff}$ in our scenario which provides a complementary cosmological probe for the verifiability of the model under consideration. 

\section{Further constraint}\label{sec::somecon}

Beyond the constraints already discussed, there are additional constraints on various model parameters arising from cosmological, experimental, and phenomenological considerations which we discuss below. \\

{\bf Higgs Invisible Decay}: 
The potential term $\frac{\lambda_{h \eta}}{2}(\Phi^\dagger \Phi)\eta^2$  and $\frac{\lambda_{h S}}{2}(\Phi^\dagger \Phi)S^2$ presented in Eq.~\ref{eqn:potential} introduces two additional decay channels for the Higgs as $S$ and $\eta$ are lighter than SM Higgs. Such decays and hence the corresponding couplings are constrained from the observation of the invisible Higgs decay. Considering the current constraint on the invisible Higgs decay branching fraction at $14.5\%$ \cite{ATLAS:2022yvh}, these couplings $\lambda_{h\eta}$ and $\lambda_{hs}$ are bounded by an upper limit of $10^{-2}$. \\

{\bf BBN Constraints}: 
As $\eta$ breaks the $Z_4$ symmetry and acquires vev which is necessary for Dirac neutrino mass generation, it mixes with the SM Higgs as already discussed in section~\ref{sec::model}, it can decay into SM charged fermions through mixing with the SM Higgs. If such decays occur after the BBN epoch, then it can alter the success of BBN predictions. 
To adhere to the BBN bound, it is crucial for $\eta$ to decay into SM particles before the onset of BBN. Specifically, the lifetime $\tau_\eta$ of $\eta$ must be shorter than $\tau_{BBN}$ and this imposes a lower limit on this mixing angle. The decay width of $\eta$ is given by
\begin{equation}
    \Gamma_\eta = \frac{M_{\eta}m_l^2 \sin\theta^2}{8 \pi v^2} \left( 1- \frac{4 m_l^2}{M_\eta^2}\right)^{3/2}
\end{equation}

\begin{figure}[h]
    \centering
    \includegraphics[scale=0.45]{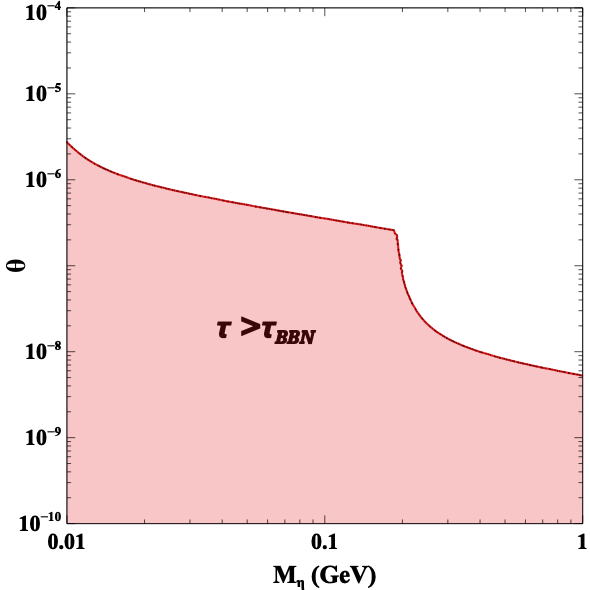}
    \caption{BBN constraint on $\eta-h$ mixing angle $\theta$ and mass of $\eta$.}
    \label{fig::lifetime}
\end{figure}

This BBN constraint on $M_\eta$ and $\theta$, is depicted in Fig.~\ref{fig::lifetime}. By setting $M_\eta =0.1$ GeV, one can obtain the lifetime of $\eta$ to be smaller than $\tau_{BBN}$ for a typical value of $\theta =10^{-6}$.\\


\section{Summary and Conclusion}\label{sec::conclusion}
In this paper, we introduce a compelling model that combines self-interacting dark matter and Dirac mass for neutrinos, utilizing a discrete $Z_4$ flavor symmetry known as 'Lepton quarticity'.  The Dirac neutrino mass is generated at the tree level, incorporating three right-chiral neutrinos ($\nu_R$), along with vector-like fermions ($N$) and a singlet scalar ($\eta$). Another $Z'_4$ symmetry is employed to prevent the direct coupling between $\nu_L$ and $\nu_R$ and to avoid undesirable couplings that may render the DM unstable. The self-interaction of DM is facilitated by a light scalar $S$. We emphasize that achieving the correct relic density of SIDM can be accomplished through the thermal freeze-out mechanism by tuning the mediator mass to a higher value in the early Universe. This adjustment addresses the issue of under-abundance resulting from excessive annihilation to $S$. Subsequently, the mass of $S$ can decrease to its present value after a phase transition that occurs well after the establishment of the DM relic density. 

We delineate two distinct cases based on the thermalisation criteria of $\nu_R$, which can yield intriguing implications from the perspective of the effective neutrino species ($\Delta N_{\rm eff}$). While direct dark matter search experiments do not impose stringent restrictions on the model parameters, $\Delta N_{\rm eff}$ offers an additional cosmological probe for the model.


\acknowledgments
SM acknowledges the support from the National Research Foundation of Korea grant no. 2022R1A2C1005050.
The work of NS is supported by the Department of Atomic Energy-Board of Research in Nuclear Sciences, Government of India (Ref. Number: 58/14/15/2021- BRNS/37220). We acknowledge Dibyendu Nanda for useful discussions.

 \section*{}
 \appendix\label{appendix}

\section{Mass Diagonalization}\label{app::neutrino mass}
The mass matrix we are dealing with here is a $6 \times 6$ matrix given by Eq.~\ref{eqn::6by6massmatrix}. To diagonalize it, we will need two unitary $6 \times 6$ matrices. However, this $6 \times 6$ matrix can be broken down into two parts. The first part will put the matrix in block diagonal form, while the second part will individually put each block matrix in their diagonal form. We take an ansatz of this unitary matrix in the form of $U=e^{iB}V$ where
\begin{equation}
    B =
    \begin{pmatrix}
        0 & S \\
        S^{\dagger} & 0 
    \end{pmatrix},~
    V =
    \begin{pmatrix}
        V_l & 0 \\
        0 & V_h 
    \end{pmatrix}
\end{equation}
where $S$ is a complex matrix, which we will take it to be $\mathcal{O}(\epsilon)$ for the perturbative approach, while $V_l$ and $V_h$ are both of order $\mathcal{O}(1)$. The subscripts $l$ and $h$ ({\it not to be confused with the Higgs field}) are written so as to remind us that $V_l$ diagonalizes the 'light' neutrino mass matrix while $V_h$ does the same for the 'heavy' $N$ mass matrix.  Then, the mass diagonalization proceeds as follows

\begin{equation}\label{app::diagonalization}
    U_L^{\dagger} M_{\nu N} U_R=V_L^{\dagger} e^{-iB_L}~M_{\nu N}~e^{iB_R}V_R
\end{equation}
Considering a small parameter $\epsilon \sim \mathcal{O}(\bold{m}/M_N)$ , $\mathcal{O}(\bold{m'}/M_N)$, such that $S \sim O(\epsilon)$. Then we can expand the $U$ matrix as 
\begin{equation}
    \begin{split}
        U =& e^{iB}.V\\
        \approx& (I + iB).V\\
        =& 
        \begin{pmatrix}
           I_3  & iS \\
           iS^\dagger & I_3
        \end{pmatrix}.
        \begin{pmatrix}
           V_l  & 0 \\
           0 & V_h
        \end{pmatrix}\\
        =& 
        \begin{pmatrix}
           V_l  & iSV_h \\
           iS^\dagger V_l & V_h
        \end{pmatrix}
    \end{split}
\end{equation}
so that \ref{app::diagonalization} becomes
\begin{equation}
    \begin{split}
        &V_L^\dagger e^{-iB_L}M_{\nu N}e^{iB_R} V_R = 
        \begin{pmatrix}
            \hat{M_\nu} & 0\\
            0 & \hat{M_N}
        \end{pmatrix}\\
        &M_{\nu N}e^{iB_R} V_R = e^{iB_L} V_L.
        \begin{pmatrix}
            \hat{M_\nu} & 0\\
            0 & \hat{M_N}
        \end{pmatrix}\\      
    \end{split}
\end{equation}
Writing the above equation in matrix form we get,
\begin{equation}
    \begin{split}
        \begin{pmatrix}
            0 & \bold{m}\\
            \bold{m'} & M_N
        \end{pmatrix}.
        &
        \begin{pmatrix}
            V_{lR} & iS_R V_{hR}\\
            iS_R^\dagger V_{lR} & V_{hR}
        \end{pmatrix}\\
         =&  \begin{pmatrix}
            V_{lL} & iS_L V_{hL}\\
            iS_L^\dagger V_{lL} & V_{hL}
            \end{pmatrix}.
         \begin{pmatrix}
            \hat{M_\nu} & 0\\
            0 & \hat{M_N}
        \end{pmatrix}
    \end{split}
\end{equation}
where the diagonal matrices are denoted with a {\it hat}~($~\hat{}~$) symbol, e.g. $\hat{M_\nu}=$diag$(M_{\nu1},M_{\nu2},M_{\nu3})$. Upon comparing the ($12$) component on both sides, we have
\begin{equation}\label{app::SL}
    \begin{split}
        &\bold{m} V_{hR} = i\, S_L V_{hL}\hat{M_N}\\
        &\implies S_L = -i\, \bold{m} V_{hR}\hat{M_N^{-1}} V_{hL}^\dagger\\
        &\implies S_L = -i\, \bold{m} M_N^{-1} \approx O(\epsilon)
    \end{split}
\end{equation}
where our assumption of $S \sim O(\epsilon)$ is justified. Similarly, by comparing the ($21$) components, one can obtain a similar expression for $S_R$ as follows:
\begin{equation}\label{app::SR}
    \begin{split}
        &\bold{m'} V_{lR} + i M_N S_R^\dagger V_{lR} = i\, S_L^\dagger V_{lL}\hat{M_\nu}\\
        &\implies \bold{m'} +i M_N S_R^\dagger = i\, S_L^\dagger V_{lL}\hat{M_\nu}V_{lR}^\dagger\\
        &\implies \bold{m'} + i M_N S_R^\dagger \approx  0 \\
        &\implies S_{R}^\dagger = i M_N^{-1} \bold{m'} \approx O(\epsilon)
    \end{split}
\end{equation}
Here, the right-hand side ($RHS$) of the second line in the above expression is neglected since it is of the order of the neutrino mass, which is extremely small compared to $M_N$ appearing on the left-hand side ($LHS$). Now, using \ref{app::SR}, the ($11$) component can be simplified as follows:
\begin{equation}
    \begin{split}
        &i \bold{m} S_R^\dagger V_{lR} = V_{lL} \hat{M_\nu}\\
        &\implies i \bold{m} S_R^\dagger = V_{lL} \hat{M_\nu}V_{lR}^\dagger =M_{\nu}\\
        &\implies M_{\nu} = i \bold{m} \left( i M_N^{-1} \bold{m'} \right)\\
        &\implies M_{\nu} = - \bold{m} M_N^{-1} \bold{m'}
    \end{split}
\end{equation}

\section{Loop Function}\label{app::loop_function}
Since $S$ does not mix with the Higgs, DM nucleon scattering can happen through a one-loop process, as shown in Fig. \ref{fig::detection}. The loop consists of one heavy fermion line along with two scalar lines. Considering the interaction with the light quarks in the external line, the effective interaction term between DM and each $q$ gives rise to the following
effective Lagrangian:
\begin{equation}
    \mathcal{L}_{eff} \supset \sum_{q=\{u,d,s\}} C_q^{tri} ~ m_q\overline{\chi}\chi \overline{q}q
\end{equation}
with the effective coupling given by
\begin{equation}\label{eqn::eff_coupling}
    C_q^{tri}=-\frac{1}{m_h^2}\frac{1}{v}~C_{H\chi\chi}
\end{equation}
The expression for the effective $H\overline{\chi}\chi$ coupling coefficient is calculated as
\begin{equation*}
    C_{H\chi\chi}=\frac{-M_\chi}{16\pi^2}\frac{v\lambda_{hs}}{2}y_{\chi}^2 \left[ \frac{\partial}{\partial p^2}B_0(p^2,M_S^2,M_\chi^2) \right]_{p^2=M_\chi^2}
\end{equation*}
The expression for the $B$ function and their derivative can be found in \cite{Abe:2018emu}. Here, we need only the derivative function and its expression is given by

\begin{eqnarray*}
    \frac{\partial B_0(p^2,M_S^2,M_\chi^2)}{\partial p^2} &=& \int_0^1 dx \frac{x(1-x)}{M_S^2x+M_\chi^2x-p^2x(1-x)}\\
         &=& \frac{-M_\chi^2 + (M_\chi^2+M_S^2)ln(\frac{M_\chi^2+M_S^2}{M_S^2})}{M_\chi^4}\\
         &=&\frac{-1+(1+(\frac{M_\chi}{M_S})^{-2})ln(1+\frac{M_\chi^2}{M_S^2})}{M_\chi^2}\\
        &=& M_\chi^{-2}L(M_\chi^2/M_S^2)
\end{eqnarray*}
where the loop function $L(x)$ is defined by 

\begin{equation}\label{app::loop}
        L(x)  = (1+x^{-1})ln(1+x)-1
\end{equation}
Then, the $H\overline{\chi}\chi$ coupling takes the form

\begin{eqnarray*}
    C_{H\chi\chi} = -\frac{y_{\chi}^2}{16\pi^2 M_\chi} \frac{v\lambda_{hs}}{2}~L(M_\chi^2/M_S^2)
\end{eqnarray*}
so that the effective coupling in \ref{eqn::eff_coupling} becomes
\begin{equation}
    C_q^{tri}=\frac{y_{\chi}^2}{16\pi^2} \frac{1}{m_h^2 M_\chi}\left[ \frac{\lambda_{hs}}{2}L(M_\chi^2/M_S^2) \right]m_q
\end{equation}\\

\section{Low energy cross-sections relevant for the self-interactions of dark matter}\label{app::SIDM cross-sections}
The non-relativistic DM self-scattering can be well understood in terms of the attractive Yukawa potential
 \begin{equation}
     V(r)=\frac{y_{\chi}^2}{4\pi r}e^{-M_{S}r}
 \end{equation}
 To capture the relevant physics of forward scattering, the transfer cross-section is defined as
 \begin{equation*}
     \sigma_T=\int d\Omega (1-cos\theta)\frac{d\sigma}{d\Omega}
 \end{equation*}
 In the Born limit, ${y_{\chi}}^2 M_{DM}/(4\pi {M_S})\ll 1$,
     \begin{equation*}
     \sigma^{\rm Born}_T=\frac{{y_{\chi}}^4}{2\pi {M_{DM}^2} v^4} \left[ log\left( 1+\frac{M^2_{DM} v^2}{M_S^2} \right)-\frac{M_{DM}^2 v^2}{M_S^2+M_{DM}^2 v^2} \right]
 \end{equation*}
 Outside the Born limit, where   ${y_{\chi}}^2 M_{DM}/(4\pi {M_S})\geq 1$, there can be two different regions: classical regime and resonance regime. In the classical regime (${M_{DM}v}/{M_S}\geq 1$), solution for an attractive potential is given by \cite{Tulin:2012wi, Tulin:2013teo, PhysRevLett.90.225002}
     \begin{equation*}
   \sigma^{\rm class.}_T=\begin{cases}
     \frac{4\pi}{M_S^2}\beta^2 ln(1+\beta^{-1}) & \textbf{$\beta > 1$}\\
     \frac{8\pi}{M_S^2}\left[ {\beta^2/(1+1.5\beta^{1.65})} \right] & \textbf{$10^{-1} < \beta \leq 10^3$} \\
     \frac{\pi}{M_S^2}\left[ ln\beta +1-1/2 ln^{-1}{\beta}\right]^2 & \textbf{$\beta \geq 10^3$}
   \end{cases}
 \end{equation*}
 where $\beta=\frac{2y_{\chi}^2{M_{S}}}{4\pi {M_{DM}}v^2}$. 

 Finally in the resonance region (${M_{DM}v}/{M_S}\leq 1$), no analytical formula for $\sigma_T$ is available. So approximating the Yukawa potential by Hulthen potential $\left(V(r)=\pm \frac{{y_{\chi}}^2}{4\pi}\frac{\delta e^{-\delta r}}{1-e^{-\delta r}}\right)$, the transfer cross-section is obtained to be: 
     \begin{equation*}
         \sigma_T^{Hulthen}=\frac{16\pi \sin^2\delta_0}{M_{DM}^2v^2}
     \end{equation*}
 where $l=0$ phase shift $\delta_0$ is given by:
 $$ \delta_0=Arg \left[ \frac{i\Gamma(iM_{DM}v/kM_S)}{\Gamma(\lambda_+)\Gamma(\lambda_-)} \right]$$
 {\rm with}
 $$\lambda_{\pm}=1+\frac{iM_{DM}v}{2kM_S}\pm \sqrt{\frac{{y_{\chi}}^2 M_{DM}}{4\pi kM_S}-\frac{M_{DM}^2{v^2}}{4k^2 {M^2_S}}}$$
 and $k \approx 1.6$ is a dimensionless number.

	\twocolumngrid
	\bibliographystyle{apsrev}

\end{document}